\documentstyle[11pt,amssymb,epsfig]{article}

\begin{document}
\newcommand{\be}{\begin{equation}}
\newcommand{\ee}{\end{equation}}
\newcommand{\ba}{\begin{eqnarray}}
\newcommand{\ea}{\end{eqnarray}}
\newcommand{\no}{\nonumber \\}
\newcommand{\gsim}{\mathrel{\hbox{\rlap{\lower.55ex \hbox {$\sim$}}
                   \kern-.3em \raise.4ex \hbox{$>$}}}}
\newcommand{\lsim}{\mathrel{\hbox{\rlap{\lower.55ex \hbox {$\sim$}}
                   \kern-.3em \raise.4ex \hbox{$<$}}}}
\def\be{\begin{eqnarray}}
\def\ee{\end{eqnarray}}
\def\bea{\be}
\def\eea{\ee}
\newcommand{\e}{{\mbox{e}}}
\def\del{\partial}
\def\vr{{\vec r}}
\def\vk{{\vec k}}
\def\vq{{\vec q}}
\def\vp{{\vec p}}
\def\vP{{\vec P}}
\def\vt{{\vec \tau}}
\def\vs{{\vec \sigma}}
\def\vJ{{\vec J}}
\def\vB{{\vec B}}
\def\hatr{{\hat r}}
\def\hatk{{\hat k}}
\def\roughly#1{\mathrel{\raise.3ex\hbox{$#1$\kern-.75em%
\lower1ex\hbox{$\sim$}}}}
\def\lsim{\roughly<}
\def\gsim{\roughly>}
\def\fm{{\mbox{fm}}}
\def\vx{{\vec x}}
\def\vy{{\vec y}}
\def\({\left(}
\def\){\right)}
\def\[{\left[}
\def\]{\right]}
\def\EM{{\rm EM}}
\def\barp{{\bar p}}
\def\zz{{z \bar z}}
\def\mus{{\cal M}_s}
\def\abs#1{{\left| #1 \right|}}
\def\ve{{\vec \epsilon}}
\def\nlo#1{{\mbox{N$^{#1}$LO}}}
\def\MS{{\mbox{M1V}}}
\def\mut{{\mbox{M1S}}}
\def\Qt{{\mbox{E2S}}}
\def\rM{{\cal R}_{\rm M1}}\def\rE{{\cal R}_{\rm E2}}
\def\la{{\Big<}}
\def\ra{{\Big>}}
\def\lsim{\mathrel{\rlap{\lower3pt\hbox{\hskip1pt$\sim$}}
     \raise1pt\hbox{$<$}}} 
\def\gsim{\mathrel{\rlap{\lower3pt\hbox{\hskip1pt$\sim$}}
     \raise1pt\hbox{$>$}}} 
\def\N{${\cal N}\,\,$}

\def\ka{{\kappa}}
\def\lam{{\lambda}}
\def\dlt{{\delta}}
\def\sig{{\sigma}}
\def\eps{{\epsilon}}
\def\wed{{\wedge}}

\def\J#1#2#3#4{ {#1} {\bf #2} (#4) {#3}. }
\def\PRL{Phys. Rev. Lett.}
\def\PL{Phys. Lett.}
\def\PLB{Phys. Lett. B}
\def\NP{Nucl. Phys.}
\def\NPA{Nucl. Phys. A}
\def\NPB{Nucl. Phys. B}
\def\PR{Phys. Rev.}
\def\PRC{Phys. Rev. C}

\renewcommand{\thefootnote}{\arabic{footnote}}
\setcounter{footnote}{0}

\vskip 0.4cm \hfill { }
 \hfill {\today} \vskip 1cm

\begin{center}
{\LARGE\bf Grazing Collisions of Gravitational Shock Waves\\
 and Entropy 
Production in Heavy Ion Collision }
\date{\today}

\vskip 1cm {\large Shu
Lin\footnote{E-mail:slin@grad.physics.sunysb.edu},
and Edward
Shuryak\footnote{E-mail:shuryak@tonic.physics.sunysb.edu}
 }


\end{center}

\vskip 0.5cm

\begin{center}

{\it Department of Physics and Astronomy, SUNY Stony-Brook,
NY 11794}

\end{center}

\vskip 0.5cm

\begin{abstract}

AdS/CFT correspondence is now widely used for study of strongly coupled
plasmas, such as produced in ultrarelativistic heavy ion collisions at RHIC.
While properties of equilibrated plasma and small deviations from equilibrium
are by now reasonably well understood, its initial formation and thermal equilibration
is much more challenging issue which remains to be studied. In the dual gravity
language, these problems are related to formation of bulk black holes, and trapped
surfaces we study in this work is a way to estimate the properties (temperature and entropy)
of such black hole. Extending the work by Gubser et al, we find numerically
trapped surfaces for non-central collision of shock waves with different
energies. We observe a critical impact parameter, beyond which the
trapped surface does not exist:  and we argue
that there are  experimental indications for similar critical
impact parameter in real collisions. 
We also present a simple solvable example of shock wave collision:
wall-on-wall collision. The applicability of this approach to
heavy ion collision is critically discussed.

\end{abstract}

\newpage

\renewcommand{\thefootnote}{\#\arabic{footnote}}
\setcounter{footnote}{0}

\section{Introduction}

The Quark Gluon Plasma(QGP) produced in Relativistic Heavy Ion Collider(RHIC)
at Brookhaven National Lab is believed to be strongly coupled \cite{SZ12}   as evidenced by its rapid equilibration, strong collective flows well reproduced by
hydrodynamics, and strong jet quenching.
 Applications of
AdS/CFT correspondence \cite{adscft} to strongly coupled QGP has generated many fundamental results
\cite{thermo,Son,CT,jet}
, for some further results see
\cite{Shuryak:2008eq} for a recent review. However this progress
so far has been mostly related either to equilibrium properties of the plasma, or
to kinetics/hydrodynamics close to equilibrium.

 The challenging issues related with violent initial stage of the collisions, in which the QGP 
is formed and equilibrated, producing most of the entropy,  are not yet understood.
One of them worth mentioning at the beginning is the strikingly different views on
 equilibration held in statistical mechanics on one hand  and in AdS/CFT-based dual gravity,
 on the other.
Statistical/kinetic approaches treat equilibration and entropy production as some gradual
deformation of particle distributions, from some initial non-thermal state toward
the final equilibrated one. In the dual gravity setting the source of temperature and entropy
are both attributed to the  gravitational horizons. Those may or may not be
produced in a collision: for example decreasing the collision energy or increasing
the impact parameter one may reach  a point at which no horizons are formed.
This implies certain singularities, or a view that switching in equilibration is
similar to a phase transition rather than a gradual deformation.

Formation of a black hole in a collision, which is then is falling toward the AdS center, was
first considered in \cite{SSZ}, with a spherical black hole. Janik and Peschanski \cite{Janik2}
have proposed asymptotic (late-time) solution, corresponding to rapidity-independent (Bjorken) flow,
see \cite{SinNak,Heller:2008fg} for most recent advances along this direction.  

Grumiller and Romatschke \cite{Grumiller:2008va} tried to describe the initial stage
of high energy collisions, starting with the gravitational shock waves
of certain type. In section 5 we will
 explore the formation of horizon in similar
setup, but taking a different point of view:
 the image on the
boundary must be due to the source in bulk. This will lead to
a different and more consistent
initial conditions, as well as subsequent evolution of matter.

A perturbative treatment of the initial conditions is
discussed by Albacete, 
Kovchegov and Taliotis \cite{kovchegov}.
Other models of equilibration based on solutions to dynamical Einstein eqns include
our previous work \cite{Lin:2008rw} in which a gravitationally collapsing shell of matter
in AdS$_5$ space is considered. It sheds light on how
 formation of isotropic and homogeneous plasma may proceed through 
 a very specific ``quasiequilibrium'' stage. We calculated the spectral
densities and found they deviate from their thermal counterpart by
general oscillations. Another interesting solution describing
isotropization of plasma was proposed by Chesler and Yaffe  \cite{Chesler:2008hg} recently.

  The issue we will address in this work is formation of trapped surfaces and
  entropy production in the collision of two shock waves in AdS background.
  The  work in this direction in AdS/CFT context  had started with the paper by
 Gubser,Pufu and Yarom \cite{Gubser:2008pc}, who considered central collisions of
 bulk pointlike black holes. They had located the
(marginally) trapped surface at the collision moment .
Its area  was then used as an estimate (a lower
bound) of the entropy produced in heavy ion collision. 
In the limit of very large collision energy $E$ they found that the entropy grows as $E^{2/3}$.
In section \ref{realistic} we will critically discuss how realistic are these results.

In this work we extend their work in two directions.
One is the obvious extension to collision of shock waves with nonzero
impact parameter. We find interesting  critical phenomenon,
 analogous to shock wave collision in Minkowski background
\cite{giddings,nambu,veneziano}: beyond certain impact parameter, 
the trapped surface disappears and black hole formation 
does not happen.
The other direction deals with
 a much simpler case of wall-on-wall collision, which was in a way 
overlooked before.

\section{Shock Waves Collision and Trapped Surface}\label{review}

It is useful to review the main steps of \cite{Gubser:2008pc} first. 
The AdS background can be written as:

\be
ds^2=L^2\frac{-dudv+(dx^1)^2+(dx^2)^2+dz^2}{z^2}
\ee

where $u=t-x^3$ and $v=t+x^3$. $x^3$ is longitudinal coordinate and $x^1,x^2$
are transverse coordinates.

The shock wave moving in $+x^3$ direction is given by:

\be\label{shock_metric}
ds^2=L^2\frac{-dudv+(dx^1)^2+(dx^2)^2+dz^2}{z^2}+L\frac{\Phi(x^1,x^2,z)}{z}\dlt(u)du^2
\ee

with $\Phi(x^1,x^2,z)$ satisfies the following equation:

\be\label{source}
\(\square-\frac{3}{L^2}\)\Phi=－16\pi G_5 J_{uu}
\ee

The 5-Dimensional source $J_{uu}$ can be arbitrary function in principle. 
$\square$ is the Laplacian in the hyperbolic space $H_3$:

\be
ds_{H3}^2=L^2\frac{(dx^1)^2+(dx^2)^2+dz^2}{z^2}
\ee

The shock wave moving in $-x^3$ direction can be obtained by the
substitution $u\leftrightarrow v$ to (\ref{shock_metric}) and (\ref{source}).

The marginally trapped surface is found from the condition of 
vanishing of expansion $\theta$\cite{hawking}. The trapped surface is
made up of two pieces: ${\cal S}={\cal S}_1\cup{\cal S}_2$. 
${\cal S}_1({\cal S}_2)$
is associated with shock wave moving in $+x^3(-x^3)$ direction before collision.
An additional condition is that the outer null normal to ${\cal S}_1$ and 
${\cal S}_2$ must be continuous at the intersection 
${\cal C}={\cal S}_1\cap{\cal S}_2$ point $u=v=0$ to avoid
delta function in the expansion.

To find out ${\cal S}_1$ associated with the first shock wave,

\be
ds^2=L^2\frac{-dudv+(dx^1)^2+(dx^2)^2+dz^2}{z^2}+L\frac{\Phi_1(x^1,x^2,z)}{z}\dlt(u)du^2
\ee

the following coordinate transformation is made to eliminate the discontinuity
in geodesics:

\be
v\rightarrow v+\frac{\Phi_1}{z}\Theta(u)
\ee

where $\Theta(u)$ is the Heaviside step function. ${\cal S}_1$ is parametrized
by:

\be
u=0,\, v=-\psi_1(x^1,x^2,z)
\ee

The expansion is defined by $\theta=h^{\mu\nu}\nabla_\mu l_\nu$, with $l_\nu$ the
outer null normal to ${\cal S}_1$. $h^{\mu\nu}$ is the induced metric. It can
be constructed from three spacelike unit vectors $w_1^\mu,\, w_2^\mu,\, w_3^\mu$,
which are normal to ${\cal S}_1$:

\be
h^{\mu\nu}=w_1^\mu w_1^\nu+w_2^\mu w_2^\nu+w_3^\mu w_3^\nu
\ee

The vanishing of expansion gives the equation:

\be\label{eom1}
\(\square-\frac{3}{L^2}\)(\Psi_1-\Phi_1)=0
\ee

with $\Psi_1(x^1,x^2,z)=\frac{L}{z}\psi_1(x^1,x^2,z)$.

The vanishing of expansion on ${\cal S}_2$ associated with the
second shock wave can be worked out similarly:

\be\label{eom2}
\(\square-\frac{3}{L^2}\)(\Psi_2-\Phi_2)=0
\ee

At the intersection ${\cal C}={\cal S}_1\cap{\cal S}_2$, ${\cal S}_1$ and
${\cal S}_2$ coincide, therefore $\Psi_1(x^1,x^2,z)=\Psi_2(x^1,x^2,z)=0$.
The continuity of outer null normal can be guaranteed by
$\nabla\Psi_1\cdot\nabla\Psi_2=4$.

In summary, the aim of finding marginally trapped surface becomes the
following unusual boundary value problem:

\be\label{bvp}
&&\(\square-\frac{3}{L^2}\)(\Psi_1-\Phi_1)=0 \no
&&\(\square-\frac{3}{L^2}\)(\Psi_2-\Phi_2)=0 \no
&&\Psi_1\vert_{\cal C}=\Psi_2\vert_{\cal C}=0
\ee

The boundary ${\cal C}$ should be chosen to satisfy the constraint:

\be\label{constraint}
\nabla\Psi_1\cdot\nabla\Psi_2\vert_{\cal C}=4
\ee

Note (\ref{bvp}) and (\ref{constraint}) are written in the form of scalar 
equation, invariant under coordinate transformation.
For central collision, the source $J_{uu}$ are identical for two shock waves.
In \cite{Gubser:2008pc}, they are chosen to be

\be\label{point}
J_{uu}=E\dlt(u)\dlt(z-L)\dlt(x^1)\dlt(x^2)
\ee

The solution of $\Phi$ corresponds to this source give rises to the following
stress tensor on the boundary field theory:

\be
T_{uu}=\frac{L^2}{4\pi G_5}\lim_{z\rightarrow 0}
\frac{\Phi(x^1,x^2,z)\dlt(u)}{z^3}=\frac{2L^4E}{\pi(L^2+(x^1)^2+(x^2)^2)^3}\dlt(u)
\ee

The special source (\ref{point}) preserves an $O(3)$ symmetry in $H_3$,
which is manifest in the following coordinate system:

\be\label{spherical}
ds_{H3}^2=\frac{dr^2}{1+r^2/L^2}+r^2\(d\theta^2+\sin^2\theta d\phi^2\)
\ee

with the point source sitting at $r=0$. We will elaborate the symmetry
later in the context of non-central collision.

The $O(3)$ symmetry helps to solve (\ref{bvp}) analytically. 
The area of the trapped surface can be calculated and give a lower bound to
the entropy produced in the collision of shock wave, assuming the area theorem
holds in AdS background.

For non-central collision, the situation is complicated by the loss of $O(3)$
symmetry. In Minkowski background, the problem of non-central collision of
point shock waves in $D=4$ was solved beautifully 
in \cite{giddings} by conformal
transformation. In $D>4$, it was solved numerically in \cite{nambu}.
In all cases, a critical impact parameter was found, beyond which the trapped
surface seized to exist.

In the next section, we will cast (\ref{bvp}) into an integral equation,
which allows us to solve (\ref{bvp}) numerically.

\section{Calculation of the Trapped Surface}

Note (\ref{bvp}) resembles the electrostatic problem in flat space, with $\Psi$
being the electric potential. We are familiar with the fact that the electric
potential can be expressed as an integral of surface charge density. We want to
see if this can be achieved in AdS space.

Let us start with the electrostatic problem in flat space. Consider
the following electrostatic problem, which is similar to (\ref{bvp}):

\be
\label{electrostatic}
&&\nabla^2\Psi_i(x)=\nabla^2\Phi_i(x) \\
&&\Psi_i(x)\vert_{\cal C}=0 \\
\label{constr_flat}
&&\nabla\Psi_1\cdot\nabla\Psi_2=4
\ee

where $i=1,2$, $\nabla^2$ is the Laplacian in flat space. $\Psi_i$ is the
electric potential corresponding to the source $\nabla^2\Phi_i$, placed
inside an empty chamber with conducting boundary ${\cal C}$. The boundary
should be chosen properly such that the constraint (\ref{constr_flat}) is also 
satisfied. 

We want to express the electric potential by an integral of the surface
charge density. This can be done with the help of the free boundary 
Green's function defined as the solution to:

\be\label{green_flat}
\nabla^2G(x,x')=\dlt^{(3)}(\vec{x}-\vec{x}')
\ee

with the solution given by:

\be
G(x,x')=-\frac{1}{4\pi}\frac{1}{\vert\vec{x}-\vec{x}'\vert}
\ee

Take (\ref{electrostatic}) multiplied by $G(x,x')$ minus (\ref{green_flat})
multiplied by $\Psi_i(x)$, and then integrate over the space inside ${\cal C}$,
we obtain:

\be
&&\int d^3x\(G(x,x')\nabla^2\Psi_i(x)-\Psi_i(x)\nabla^2G(x,x')\)=
\int d^3xG(x,x')\nabla^2\Phi_i(x)-\Psi_i(x') \\
\label{surface_flat}
&&\int d{\vec S}\cdot\(G(x,x'){\vec \nabla}\Psi_i(x)-\Psi_i(x){\vec \nabla} G(x,x')\)=
\int d^3xG(x,x')\nabla^2\Phi_i(x)-\Psi_i(x')
\ee

Denote $B_i(x)=-\frac{\del\Psi_i(x)}{\del n}$(the magnitude of electric field
on the boundary) and note $\Psi_i(x)$ vanishes
on the boundary. With $x'$ taken on the boundary ${\cal C}$, 
(\ref{surface_flat}) evaluates to:

\be\label{integral_flat}
\int dS G(x,x')B_i(x)=\int d^3xG(x,x')\nabla^2\Phi_i(x)
\ee

The constraint (\ref{constr_flat}) is simply $B_1(x)B_2(x)=4$. 
We have cast a problem in the volume into a problem on its boundary ${\cal C}$.
(\ref{integral_flat}) is a Fredholm integral equation of the first kind. We
can use the following method to solve (\ref{electrostatic}): Starting with
 some trial shape of ${\cal C}$, we can solve 
(\ref{surface_flat}) to obtain $B_i(x)$ and check if (\ref{constr_flat})
is satisfied. We can use iteration to tune the trial shape 
until (\ref{constr_flat}) is satisfied.

Now we hope to apply similar method to the problem of trapped surface, the
difference being the space is $H_3$ instead of flat.

As in case of electrostatic problem, we will keep using Green's function in
AdS, defined as the solution to the following:

\be
\(\square-\frac{3}{L^2}\)G(x,x')=\frac{1}{\sqrt{g}}\dlt^{(3)}(\vec{x}-\vec{x}')
\ee

where $g$ is the metric of $H_3$.

The Green's function was solved in \cite{maldacena,danielsson}. We quote the
result here with $L$ dependence restored.

\be\label{green}
&&G(x,x')=-\frac{1}{4\pi L}\frac{e^{2u}}{\sinh u} \no
&&\cosh u=1+\frac{(z-z')^2+(\vec{x}_\perp-\vec{x'}_\perp)^2}{2zz'}
\ee

where $u$ is the invariant distance in $H_3(AdS_3)$.

It also proves useful to note another relation:

\be\label{stokes}
&&\int_M\square f\sqrt{g} d^3x \no
&&=\int_M\frac{1}{\sqrt{g}}\del_\mu\(\sqrt{g}g^{\mu\nu}\del_\nu f\)\sqrt{g}\frac{1}{3!}\eps_{\sig\rho\lam}dx^\sig\wed dx^\rho\wed dx^\lam \no
&&=\int_M d(\sqrt{g}g^{\mu\nu}\del_\nu f\eps_{\mu\rho\lam}\frac{1}{2!}dx^\rho\wed dx^\lam)
\ee

where $\overline{dx}^\nu=g^{\mu\nu}\sqrt{g}\eps_{\mu\rho\lam}\frac{1}{2!}dx^\rho\wed dx^\lam$. $M$ is taken to be the manifold in $H_3$ bounded by ${\cal C}$, the metric $g$ refers to $H_3$. $f$ is arbitrary function of $x$.

With (\ref{stokes}) and (\ref{green}) at hand, we are ready to proceed:

\be\label{eqns_ads}
\left\{\begin{array}{l}
\(\square-\frac{3}{L^2}\)\Psi_i(x)=\(\square-\frac{3}{L^2}\)\Phi_i(x) \\
\(\square-\frac{3}{L^2}\)G(x,x')=\frac{1}{\sqrt{g}}\dlt^{(3)}(\vec{x}-\vec{x}')
\end{array}
\right.
\ee

with $i=1,2$. All the derivatives are with respect to $x$.
The first line of (\ref{eqns_ads}) multiplied by $G(x,x')$ minus the second line of
(\ref{eqns_ads}) multiplied by $\Psi_i(x)$, then integrate over $M$, we obtain:

\be
&&\int_{M}\(G(x,x')\(\square-\frac{3}{L^2}\)\Psi_i(x)-\Psi_i(x)\(\square-\frac{3}{L^2}\)G(x,x')\)\sqrt{g}d^3x= \no
&&\hspace{5cm}\int_{M}G(x,x')\(\square-\frac{3}{L^2}\)\Phi_i(x)\sqrt{g}d^3x-\Psi_i(x') \\
&&\int_{M}\(G(x,x')d\(\del_\nu\Psi_i(x)\overline{dx}^\nu\)-\Psi_i(x)d\(\del_\nu G(x,x')\overline{dx}^\nu\)\)= \no 
&&\hspace{5cm}\int_{M}G(x,x')\(\square-\frac{3}{L^2}\)\Phi_i(x)\sqrt{g}d^3x-\Psi_i(x') \\
&&\int_{M}\(d\(G(x,x')\del_\nu\Psi_i(x)\overline{dx}^\nu\)-d\(\Psi_i(x)\del_\nu G(x,x')\overline{dx}^\nu\)\)= \no
&&\hspace{5cm}\int_{M}G(x,x')\(\square-\frac{3}{L^2}\)\Phi_i(x)\sqrt{g}d^3x-\Psi_i(x') \\
&&\int_{\del M}\(G(x,x')\del_\nu\Psi_i(x)\overline{dx}^\nu-\Psi_i(x)\del_\nu G(x,x')\overline{dx}^\nu\)= \no
&&\hspace{5cm}\int_{M}G(x,x')\(\square-\frac{3}{L^2}\)\Phi_i(x)\sqrt{g}d^3x-\Psi_i(x')
\ee

where in the last line we have used Stokes theorem on manifold $M$.

Putting $x'$ on ${\cal C}$, we can simplify the above with 
$\Psi_i\vert_{\cal C}=0$:

\be\label{boundary}
\int_{\del M}G(x,x')\del_\nu\Psi_i(x)\overline{dx}^\nu=\int_{M}G(x,x')\(\square-\frac{3}{L^2}\)\Phi_i(x)\sqrt{g}d^3x
\ee

Furthermore, we have $\del_\nu \psi dx^\nu=0$ on ${\cal C}$ since 
$\Psi_i\vert_{\cal C}=0$. On the other hand, $n_\nu dx^\nu\vert_{\cal C}=0$, where
$n_\nu$ is the unit vector normal to the boundary ${\cal C}$.
Therefore, we may write:

\be\label{B}
\del_\nu\Psi_i=-B_i n_\nu
\ee

With the help of (\ref{B}), (\ref{boundary}) and (\ref{constraint}) 
can be further simplified to:

\be
\label{boundary_B}
&&-\int_{\del M}G(x,x')B_i(x)dS=\int_{M}G(x,x')\(\square-\frac{3}{L^2}\)\Phi_i(x)\sqrt{g}d^3x \\
\label{product}
&&B_1(x)B_2(x)=4
\ee

where $dS\equiv n_\mu\overline{dx}^\mu$ is the area element.

Before proceeding to non-central collision, we would like to reproduce the 5-D
result of \cite{Gubser:2008pc} first. Working in spherical coordinates 
(\ref{spherical}), the shape of ${\cal C}$ is parametrized by $r=\rho_0=const$.
The simplest point shock wave corresponding to 
$J_{uu}=E\dlt(u)\dlt(z-L)\dlt(x^1)\dlt(x^2)$ is given by:

\be
&&\Phi_1=\Phi_2=\frac{4G_5E}{L}\frac{1+2(r/L)^2-2r/L\sqrt{1+(r/L)^2}}{r/L}
\ee

The Green's function (\ref{green}) is invariant under coordinate transformation.
In spherical coordinate, it is given by:

\be
&&G(x,x')=-\frac{1}{4\pi L}\frac{e^{2u}}{\sinh u} \no
&&\cosh u=\sqrt{r^2/L^2+1}\sqrt{r'^2/L^2+1}-rr'/L^2
\(\cos\theta\cos\theta'+\sin\theta\sin\theta'\cos(\phi-\phi')\)
\ee

In the presence of $O(3)$ symmetry, it is sufficient to show (\ref{boundary_B})
holds for $\theta'=0$, when the integral in $\phi$ is trivial. 
On the other hand, (\ref{product}) implies $B_1=B_2=2$. As a result,
we only need to verify:

\be
&&2\pi\int_0^\pi d\theta(-2)\frac{(\cosh u-\sinh u)^2}{\sinh u}\rho_0^2\sin\theta
=\frac{(\sqrt{\rho_0^2+1}-\rho_0)^2}{\rho_0}(-4G_5 E)4\pi \no
&&\cosh u=\rho_0^2+1-\rho_0^2\cos\theta
\ee

It is not difficult to complete the integral in $\theta$, we finally arrive at
$2G_5E=\sqrt{1+\rho_0^2/L^2}\rho_0^2$, which is equivalent to (115) 
in \cite{Gubser:2008pc}.

\section{Colliding Point Shock Waves at nonzero Impact Parameter}

\subsection{Shock Waves in Spherical Coordinate}

Consider two shock waves with impact parameter $b$, given by:

\be
&&\(\square-\frac{3}{L^2}\)\Psi_1=-16\pi G_5E\dlt(u)\dlt(z-z_0)\dlt(x^1-\frac{b}{2})\dlt(x^2)\\
&&\(\square-\frac{3}{L^2}\)\Psi_2=-16\pi G_5E\dlt(u)\dlt(z-z_0)\dlt(x^1+\frac{b}{2})\dlt(x^2)\nonumber
\ee

The corresponding stress energy tensor associated with two shock waves
are given by:

\be
&&T_{uu}=\frac{2L^4E}{\pi(z_0^2+(x^1-\frac{b}{2})^2+(x^2)^2)^3}\dlt(u) \no
&&T_{vv}=\frac{2L^4E}{\pi(z_0^2+(x^1+\frac{b}{2})^2+(x^2)^2)^3}\dlt(v) \nonumber
\ee

Therefore $z_0$ characterizes the size of the nucleus. 
We will use spherical coordinates in solving (\ref{boundary_B}). In case of
central collision, when $b=0$. The shock wave center can be placed 
at the origin of spherical coordinates $r=0$. 
This is achieved by first going to global coordinates 
$Y^i(i=0,1,2,3)$:

\be\label{p2g_k}
&&Y^0=\frac{z}{2}\(k+\frac{L^2/k+kx_\perp^2}{z^2}\) \no
&&Y^3=\frac{z}{2}\(-k+\frac{L^2/k-kx_\perp^2}{z^2}\) \no
&&Y^1=L\frac{x^1}{z} \no
&&Y^2=L\frac{x^2}{z}
\ee

The global coordinates link to spherical coordinates in the following way:

\be\label{g2s}
&&Y^0=\sqrt{r^2+L^2} \no
&&Y^1=r\cos\theta \no
&&Y^2=r\sin\theta\cos\phi \no
&&Y^3=r\sin\theta\sin\phi
\ee

When $b=0$, the center of the shock waves can be put at the origin if we set
$k=\frac{L}{z_0}$.  The possibility of moving
any point to the origin reflects the maximally symmetric property of AdS
space. 
 
When $b\neq 0$, we want to place the two shock waves at opposite
positions with respect to the origin, so that the boundary of trapped surface
${\cal C}$ will have axial symmetry. 
Setting $1+\frac{b^2}{4z_0^2}=\frac{L^2}{k^2z_0^2}$, we have $Y^2=Y^3=0$ and 
$Y^1=\pm\frac{Lb}{2z_0}$. According to (\ref{g2s}), we have the shock waves at
$r=\frac{Lb}{2z_0}$, $\theta=0$ and $\theta=\pi$. The differential equation
in (\ref{boundary_B}) becomes:

\be
\(\square-\frac{3}{L^2}\)\Psi_i=
-16\pi G_5E\frac{L^3}{z_0^3}\frac{\sqrt{1+r^2/L^2}}{r^2\sin\theta}\dlt(r-r_0)
\dlt(\theta-\theta_i)\dlt(\phi)
\ee

where $r_0=\frac{Lb}{2z_0}$, $\theta_1=0$, $\theta_2=\pi$. We observe that
in spherical coordinate, the trapped surface only depends on 
$G_5E\frac{L^3}{z_0^3}$ and $r_0$. Since AdS radius is a free parameter, which
will not appear alone in the final result in dual field theory, we may set
$z_0=L$ without loss of generality. As a result we have $b=2r_0$.

\subsection{More General Shock Waves}

Before proceeding to numerical study of trapped surface, we choose to take
a moment to investigate the symmetries of the problem, which will help us
to study more general shock waves. To see this, we prefer to work in the
differential form of the problem: (\ref{bvp}) and (\ref{constraint}).

As we noticed before, (\ref{bvp}) and (\ref{constraint}) are scalar equations.
$\Psi_i$ is a scalar. It is invariant under coordinate transformations:
 $x\rightarrow\tilde{x}$, $\Psi_i(x)\rightarrow\tilde\Psi_i(\tilde{x})$
the boundary remain the same ${\cal C}\rightarrow\tilde{\cal C}$,
but takes a different functional form in
new coordinate. As a result the third line of (\ref{bvp}) and 
(\ref{constraint}) are automatically satisfied. Suppose the transformation
also preserves the form of the operator: $\square-3/L^2$, then 
$\tilde\Psi_i(\tilde{x})$ becomes another solution to (\ref{bvp}) and 
(\ref{constraint}). We will focus on transformations that leaves the center
of the shock waves on the axis of $\theta=0,\,\pi$.

To identify such a coordinate transformation, we first make a change of 
variable:

\be\label{eta}
&&r\sin\theta=t \no
&&r\cos\theta=\sqrt{L^2+t^2}\sinh\eta \nonumber
\ee

The metric of $H_3$ becomes:

\be\label{metric_eta}
ds^2=\frac{dt^2}{1+t^2/L^2}+(L^2+t^2)d\eta^2+t^2d\phi^2
\ee

The metric is $\eta$ independent, therefore the transformation: 
$\tilde{t}=t$,$\tilde{\phi}=\phi$,$\tilde{\eta}=\eta+\Delta\eta$ will not 
change the operator $\square-3/L^2$. $\tilde{t}=t$ also guarantees the
center of the shock waves remain on the axis of $\theta=0,\,\pi$.
We have obtained the desired coordinate transformation, which is just
a translation in $\eta$. It is easy to work out the corresponding
transformation in spherical coordinate:

\be\label{transform}
&&\tilde{r}\sin\tilde{\theta}=r\sin\theta=t \no
&&\tilde{r}\cos\tilde{\theta}=\sqrt{L^2+t^2}\sinh(\eta-\Delta\eta) \no
&&r\cos\theta=\sqrt{L^2+t^2}\sinh\eta
\ee

One can verify explicitly (\ref{transform}) preserves the form of 
(\ref{spherical}).
(\ref{transform}) moves the center of the shock waves from $Y^2=Y^3=0$,
$Y^1=\pm r_0$ to $Y^2=Y^3=0$, 
$Y^1=\pm r_0\cosh\Delta\eta-\sqrt{L^2+r_0^2}\sinh\Delta\eta$. This means
collision of shock waves centered at $Y^2=Y^3=0$,
$Y^1=\pm r_0\cosh\Delta\eta-\sqrt{L^2+r_0^2}\sinh\Delta\eta$ will generate
the same entropy as those centered at $Y^2=Y^3=0$,
$Y^1=\pm r_0$. This allows us to study the collision of more general 
shock waves. Let us consider the following shock waves:

\be\label{shock}
&&\(\square-\frac{3}{L^2}\)\Psi_1=-16\pi G_5E_u\dlt(u)\dlt(z-z_u)\dlt(x^1-x_u)\dlt(x^2) \no
&&\(\square-\frac{3}{L^2}\)\Psi_2=-16\pi G_5E_v\dlt(v)\dlt(z-z_v)\dlt(x^1-x_v)\dlt(x^2)
\ee

In this paper, we restrict our interest in shock waves with identical invariant
energy, defined by $E_u\frac{L^3}{z_u^3}=E_v\frac{L^3}{z_v^3}\equiv E$. This
keeps the mirror symmetry of the problem intact.
We will see the center of the shock waves can be placed at $Y^2=Y^3=0$, 
$Y^1=\pm r_0\cosh\Delta\eta-\sqrt{L^2+r_0^2}\sinh\Delta\eta$. This is
equivalent to the statement that a solution can always be found to
the following equations:

\be\label{eqns}
&&L\frac{x_u}{z_u}=r_0\cosh\Delta\eta-\sqrt{L^2+r_0^2}\sinh\Delta\eta \no
&&L\frac{x_v}{z_v}=-r_0\cosh\Delta\eta-\sqrt{L^2+r_0^2}\sinh\Delta\eta \no
&&k^2\(1+\frac{x_u^2}{z_u^2}\)=\frac{L^2}{z_u^2} \no
&&k^2\(1+\frac{x_v^2}{z_v^2}\)=\frac{L^2}{z_v^2} \no
&&x_u-x_v=\Delta x
\ee

(\ref{eqns}) can be solved easily by switching to the variable 
$\eta_0=\sinh^{-1}\frac{r_0}{L}$. A solution to (\ref{eqns}) always exists
for any given $z_u$, $z_v$ and $\Delta x$. We include the corresponding 
$r_0$ here, as it is the only relevant quantity, apart from $E$,
for entropy calculation:

\be\label{r0}
\frac{r_0}{L}=\sqrt{\frac{(z_u-z_v)^2+\Delta x^2}{4z_uz_v}}
\ee

In summary, we have shown that for shock waves (\ref{shock}) satisfying
$E_u\frac{L^3}{z_u^3}=E_v\frac{L^3}{z_v^3}\equiv E$, the entropy is only
a function of $G_5E$ and (\ref{r0}). Note $r_0$ is only a function
of invariant distance between the center of shock waves.

\subsection{Numerical Solution of Trapped Surface}

With all the simplification, we are ready to find the trapped surface for
different impact parameter. Our procedure is as follows:
Axial symmetry allows us to parametrize ${\cal C}$ by
$r=\rho(\theta)$. Integral in $\phi$ on the LHS of (\ref{boundary_B}) can be
expressed in terms of elliptic integrals. (\ref{boundary_B}) becomes essentially
1-D integral equation. We discretize the integral by $199$ points, equally
spaced in the full range of $\theta$. The integral on the LHS of 
(\ref{boundary_B}) is discretized accordingly, and the integral on the RHS can
be expressed in terms of elementary function due to the simple form of the
shock wave. We use the same sample points for $\theta'$, bringing 
(\ref{boundary_B}) into a matrix form:

\be\label{matrix}
\sum_j B(\theta_j)K(\theta_j,\theta'_i)=S(\theta'_i)
\ee

where the indices $i,j=1,\cdots,199$. $K(\theta_j,\theta'_i)$ contains the
Green's function and the induced metric. $S(\theta'_i)$ is from the RHS
integral of shock wave.

A special treatment is needed for diagonal matrix element of 
$K(\theta_j,\theta'_i)$ where $\theta_j=\theta'_i$. The explicit integrand
expressed in terms of elliptic integrals shows the it is logarithmically
divergent in $\vert\theta-\theta'\vert$, yet the integral is convergent. The integral
in this interval, represented by the diagonal matrix element, is estimated
by sampling the integrand by certain number of points in the interval. 
The sample integrand are used to extract the coefficients of terms
$\ln\vert\theta-\theta'\vert$,
$1$, $(\theta-\theta')\ln\vert\theta-\theta'\vert$ and $\theta-\theta'$ by
method of least squares. Those coefficients are finally used for calculation of
diagonal matrix elements.

The mirror symmetry of the two shock waves implies $B_2(\theta)=B_1(\pi-\theta)$.
Therefore it is sufficient to calculate one of them. Given a trial shape of
trapped surface $r=\rho(\theta)$, which is also necessarily symmetric under
$\theta\leftrightarrow\pi-\theta$, we can solve for $B(\theta)$ from 
(\ref{matrix}). We then evaluate $\Delta(\theta)=B_1(\theta)B_2(\theta)-4$ 
and tune the 
shape function accordingly. We repeat the process until (\ref{product}) is
satisfied to certain accuracy. In order to assure fast convergence, we find
it very helpful to
calculate the gradient of $\rho(\theta)$. The gradient is the matrix form
of the functional derivative: 
$\frac{\dlt\Delta[\rho(\theta)]}{\dlt\rho(\theta)}$. We parametrized ${\cal C}$
by: $\rho(\theta)=\sum_{n=1}^{M} a_n\cos{2(n-1)\theta}$, where $M$ is
a truncation number. The same 
decomposition applies to $\Delta(\theta)$:
$\Delta(\theta)=\sum_{n=1}^{M} b_n\cos{2(n-1)\theta}$. 
The gradient in this representation is given by a $M\times M$ matrix:
$\frac{\dlt b_m}{\dlt a_n}$, which again contains elliptic integrals.
For a given collision energy, we can find the boundary ${\cal C}$ until certain
critical impact parameter is reached. The critical impact parameter is
located where $\frac{\del\rho(\theta)}{\del b}$ diverges\cite{nambu}.
Empirically, the gradient $\frac{\dlt b_m}{\dlt a_n}$ converges as 
$\Delta(\theta)$ reduces in the iteration, if the impact parameter
is within the critical value. The gradient diverges as $\Delta(\theta)$ 
reduces in the iteration, if the impact parameter lie beyond the critical
value. 

\begin{figure}[t]
\includegraphics[width=0.4\textwidth]{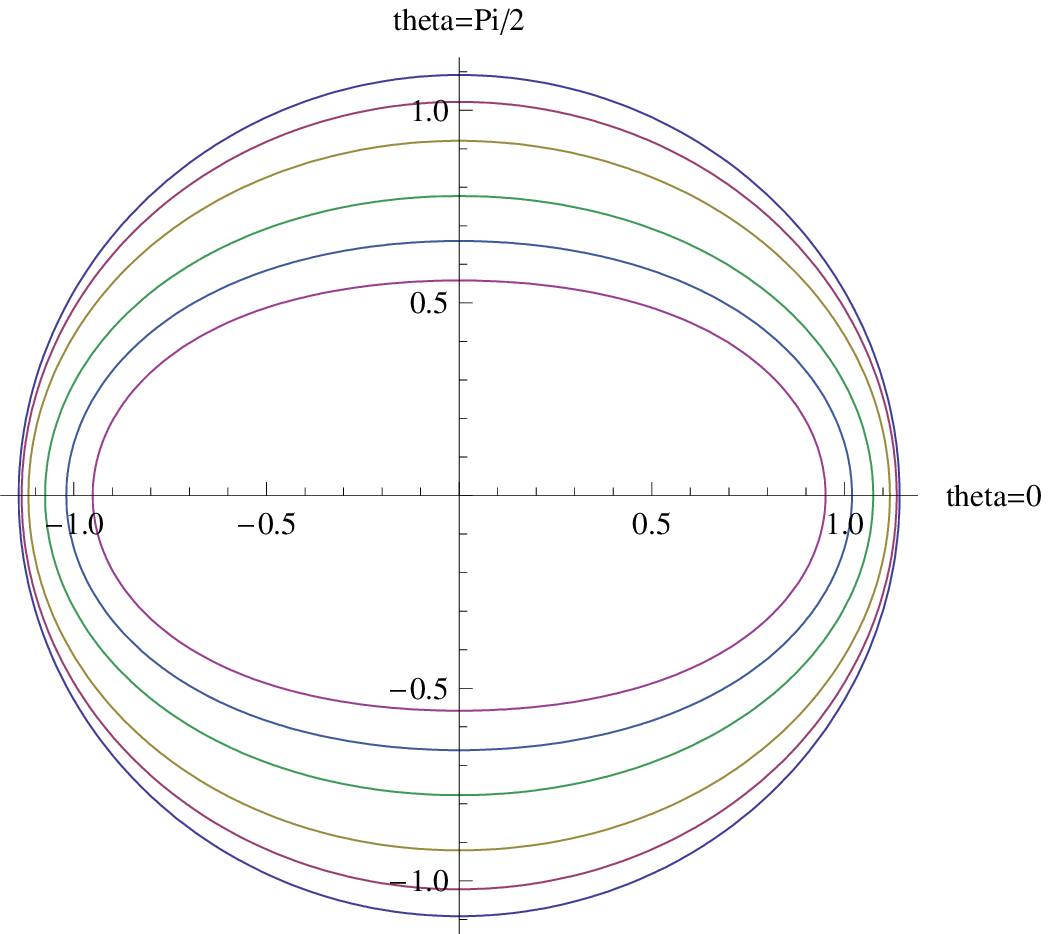}
\includegraphics[width=0.4\textwidth]{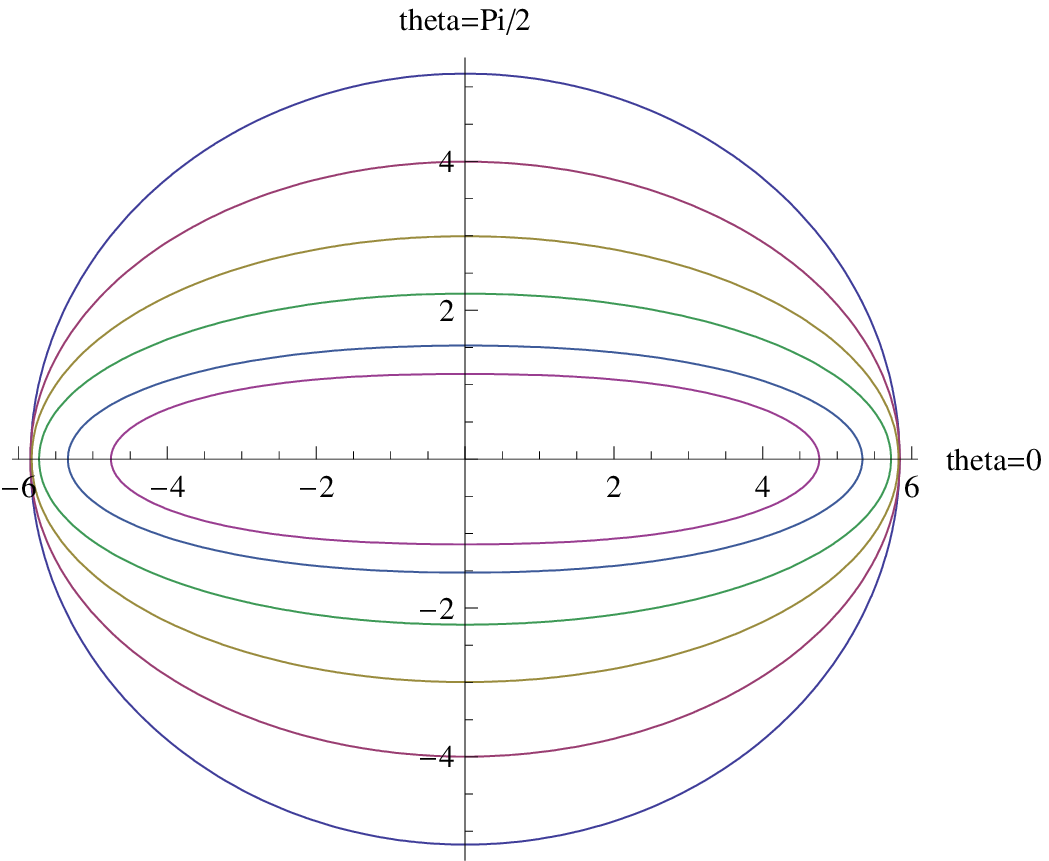}
\caption{\label{fig_shapes}(left)The shapes of ${\cal C}$
(the trapped surface at $u=v=0$) at $\frac{G_5E}{L^2}=1$.
The impact parameters used in the plot are $0.4L,\,0.6L,\,0.8L,\,
1.0L,\,1.1L,\,1.14L$ from the outer to the inner. The innermost shape
being the critical trapped surface.
(right)The shapes of ${\cal C}$
(the trapped surface at $u=v=0$) at $\frac{G_5E}{L^2}=100$.
The impact parameters used in the plot are $1.0L,\,2.0L,\,3.0L,\,
4.0L,\,5.0L,\,5.3L$ from the outer to the inner. The innermost shape
being the critical trapped surface.
 As collision energy grows, the trapped surface gets elongated
in the axis of mismatch.}
\end{figure}

Fig.\ref{fig_shapes} shows the shapes of trapped surface at
$\frac{G_5E}{L^2}=1$ and $\frac{G_5E}{L^2}=100$ for different impact parameters. 
The shapes are represented in spherical coordinate. We observe
the critical trapped surface does not scale with collision energy in spherical
coordinate. As collision energy grows, the trapped surface gets elongated
in the axis of mismatch and larger $M$ is needed to reach prescribed accuracy.

We also obtained several critical impact parameters corresponding to different
energies. The results are listed in Table.\ref{tab:crit}

\begin{table}
\caption{\label{tab:crit} critical impact parameter at different energies}
\begin{tabular}{cccccccccc}
$\frac{G_5E}{L^2}$& 0.1& 0.5& 1& 4& 9& 12& 15& 50& 100\\
\hline
$\frac{b_{c}}{L}$& 0.40& 0.86& 1.14& 1.90& 2.50& 2.74& 2.94& 4.28& 5.30\\
\end{tabular}
\end{table}

\begin{figure}[t]
\includegraphics[width=0.5\textwidth]{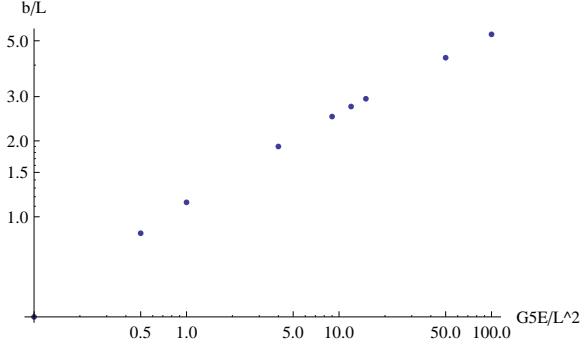}
\caption{\label{Eb}The log-log plot of critical impact parameter versus
collision energy.}
\end{figure}

Fig.\ref{Eb} shows the log-log plot of critical impact parameter versus
collision energy. It suggests a simple power law within the energy range used
in the numerical study.
The data are fitted with $\frac{b_{c}}{L}=\alpha\(\frac{G_5E}{L^2}\)^\beta$
to give:

\be\label{fit}
\alpha=1.07, \quad \beta=0.37
\ee

$b\sim E^\beta L^{1-2\beta}$, the numerical value from fitting shows the
critical impact parameter grows with collision energy and nucleus size.

The area of the trapped surface(twice the area of ${\cal C}$)
 sets a lower bound of the entropy produced,
given as follows:

\be\label{lb}
&&S_{trapped}=\frac{2A}{4G_5}=\frac{1}{2G_5}\int\sqrt{g}d^3x 
\ee

where $A$ is the area of the boundary ${\cal C}$. The prefactor is

\be
\frac{L^3}{G_5}=\frac{2N_c^2}{\pi}
\ee

\begin{figure}[t]
\includegraphics[width=0.5\textwidth]{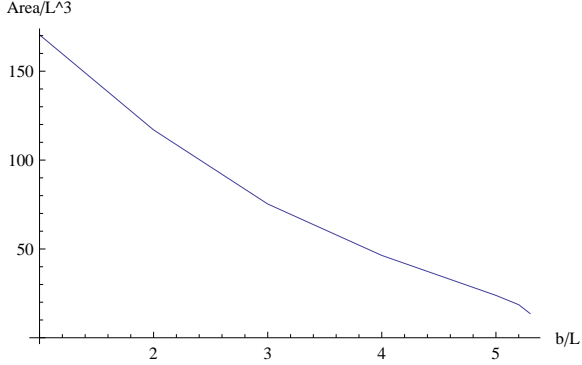}
\caption{\label{fig_entropy}The scaled entropy $2G_5S/L^3$(the area of 
${\cal C}$) as a function of the
impact parameter scaled $b/L$. The energy used is 
$\frac{G_5E}{L^2}=100$, where $\frac{L^3}{G_5}=\frac{2N_c^2}{\pi}$}
\end{figure}

We plot the lower bound of entropy in the dual field theory for
energy $\frac{G_5E}{L^2}=100$ in Fig.\ref{fig_entropy}

\section{Wall-on-wall collisions}

 In this section we address  a simpler form of the shock waves, called wall-on-wall in \cite{SSZ},
in which there is no dependence on two transverse coordinates. 
Grumiller and Romatschke 
\cite{Grumiller:2008va}  have also discussed it,
 using gravitational shock waves. The problem with their approach
is (power) growing amplitude of the shock as a function of holographic
coordinate $z$. If so, collision dynamics resembles the atmospheric turbulence in 
the sense that the largest perturbation is at the largest $z$ -- namely in the infrared modes --
cascading down toward higher momenta (UV). First of all
 this  is physically  different from the
 heavy ion collisions, in which case the initial wave function have partons well localized
near the so called  ``saturation scale'', from which
the equilibration domain (trapped surface)
 propagates both into small $z$ (UV) and large $z$ (IR)
directions.
  Second,  we think it is also inconsistent: 
 a function growing with $z$
cannot be considered as a small perturbation to
the  background metric, which is decreasing at large $z$ as $1/z^2$.
One may think that gravity near the AdS center $z=\infty$ should
never be touched, as this is the original positions of the $D_3$
branes which gave the basis to AdS/CFT correspondence 
in the first place. 

 Our
choice of the initial conditions, describing colliding walls with fixed parton
 density and thus fixed saturation scale is given by  the following source
\be
\(\square-\frac{3}{L^2}\)\Phi(z)=-16\pi \frac{G_5E}{L^2}\dlt(z-z_0)
\ee
The corresponding solution to Einstein eqn subject to the boundary condition $\Psi(z)\rightarrow 0$ as
$z\rightarrow 0$ is easily obtained:

\be
\Phi(z)=\left\{\begin{array}{l}
4\pi G_5E\frac{z^3}{z_0^4}\quad z<z_0 \\
4\pi G_5E\frac{1}{z}\quad z>z_0 
\end{array}
\right.
\ee
Note that it decreases in both direction from the original scale $z_0$: therefore
(as we will see shortly) the trapped surface has finite extensions in both directions from it.

The corresponding  stress energy tensor on the boundary (as seen by an observer living in dual gauge theory) is
\be T_{uu}=\frac{EL^2}{z_0^4}\dlt(u) \ee
The stress energy tensor is the same as that used in \cite{Grumiller:2008va,kovchegov}, and our solution converges well as $z\rightarrow 0$.
We choose to collide states with different energy therefore 
we fix $z_0$, but use different $E$.
Applying the general discussion of shock wave in Sec.\ref{review} and noting
the trapped surface only depends on $z$, we obtain:

\be\label{plane}
&&z^2\Psi_i''-z\Psi_i'-3\Psi_i=-16\pi G_5E_i\dlt(z-z_0) \\
&&\Psi_i(z_a)=\Psi_i(z_b)=0 \\
\ee

with $i=1,2$. $\Psi_1$ and $\Psi_2$ are shape functions corresponding to two
shock waves. The trapped region at $u=v=0$ is limited by the interval $z_a<z<z_b$. 
The constraint (\ref{product}) takes a simple form:

\be\label{plane_cstr}
&&\Psi_1(z_a)\Psi_2(z_a)\frac{z_a^2}{L^2}=4 \no
&&\Psi_1(z_b)\Psi_2(z_b)\frac{z_b^2}{L^2}=4
\ee

(\ref{plane}) is easily solved to give:

\be\label{plane_sol}
&&\Psi_i(z)=\left\{\begin{array}{l}
C\(\(\frac{z}{z_a}\)^3-\frac{z_a}{z}\)\quad z<z_0 \\
D\(\(\frac{z}{z_b}\)^3-\frac{z_b}{z}\)\quad z>z_0 
\end{array}
\right. \no
&&C=-4\pi G_5E_i\frac{\(\frac{z_0}{z_b}\)^3-\frac{z_b}{z_0}}{\(\frac{z_0}{z_a}\)^3z_b-\(\frac{z_0}{z_b}\)^3z_a} \no
&&D=-4\pi G_5E_i\frac{\(\frac{z_0}{z_a}\)^3-\frac{z_a}{z_0}}{\(\frac{z_0}{z_a}\)^3z_b-\(\frac{z_0}{z_b}\)^3z_a}
\ee

Plugging (\ref{plane_sol}) in (\ref{plane_cstr}), we obtain:
\be\label{cardano}
&&z_a+z_b=\frac{8\pi G_5\sqrt{E_1E_2}}{L} \\
&&\frac{(z_a+z_b)^2-3z_az_b}{(z_az_b)^3}=\frac{L^3}{z_0^4}
\ee

Note $E_1,\,E_2$ appear only in the combination $\sqrt{E_1E_2}$, This is
consistent with the picture that only the center of mass contributes to the
entropy. Recall the
the center of mass of two massless particles with energy $E_1,\,E_2$ is
 $2\sqrt{E_1E_2}$. The resulting cubic eqn (\ref{cardano}) can 
be solved by
Cardano formula, but the explicit solution is not illustrative and is not
showed here. The entropy is given by:

\be
&&S=\frac{2A}{4G_5}=\frac{\int\sqrt{g}dzd^2x_\perp}{2G_5} \no
&&s\equiv\frac{S}{\int d^2x_\perp}=\frac{L^3}{4G_5}(\frac{1}{z_a^2}-\frac{1}{z_b^2})
\ee

The leading behavior of entropy per transverse area $s$ in energy is extracted:

\be 
s\sim\frac{4L^2}{z_0^4}(\pi G_5\sqrt{E_1E_2}z_0^2)^\frac{2}{3}
\ee

The power $2/3$ is the same as point shock wave obtained in \cite{Gubser:2008pc}.
There is also an obvious lower bound of the energy for the formation of
trapped surface:

\be\label{bound}
4\pi G_5E\ge z_0
\ee

\begin{figure}[t]
\includegraphics[width=0.5\textwidth]{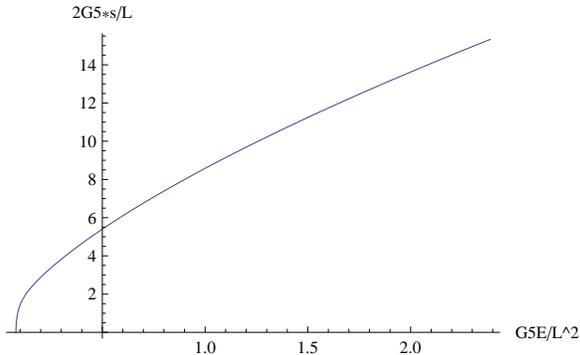}
\caption{\label{fig_plane}The scaled entropy per transverse area 
$\frac{2G_5 s}{L^2}$(the area of ${\cal C}$ per transverse area)
as a function of scaled effective colliding energy $G_5E/L^2$, where
$\frac{L^3}{G_5}=\frac{2N_c^2}{\pi}$. }
\end{figure}

The equality is reached at $z_a=z_b$, when the ${\cal C}$ has vanishing volume.
For general energy, $s$ is evaluated as a function of effective colliding energy
$E=\sqrt{E_1E_2}$. We again set $z_0=L$.
Fig.\ref{fig_plane} shows the entropy as a function of effective colliding
energy.

\section{Matching heavy ion collisions to those of gravitational shock waves}
\label{realistic}
As the reader who came to this point of the paper knows, it 
was up to this point  a methodical
work devoted to solving well-posed mathematical problems set in
the AdS/CFT  dual-gravity
framework.  However now, near the end of this work,
 we would like to address
a wider issues of applicability limits of such approach, as well as the best strategy
to use it for practical problems.

Gubser et al \cite{Gubser:2008pc} have applied the gravitational collision scenario $literally$,
selecting initial conditions at time long before nuclei collide. More specifically, they have
(i) tuned the scale $L$ or $z_0$ of the bulk colliding object to the size of the nucleus $R$ and
(ii) have used the realistic CM gamma factor of the colliding nuclei $E/m=\gamma\sim 100$.
The result of such choice is a {\em completely unrealistic} fireball produced, in spite
of a reasonable entropy. Indeed, the size of the trapped surface \cite{Gubser:2008pc} is huge, about $300 fm$,
which is very large compared to colliding nuclei.
In real heavy ion collisions the produced fireball
 has the same size as the nuclei, with the radius about $6 fm$. The initial
temperature -- as estimated by $z_{min}\sim 1/\pi T_i$ where  $z_{min}$ is the minimal distance of the 
trapped surface to the AdS boundary -- is however way too high. So, what went wrong with this straightforward approach?

The answer to this question is in fact well known:  initial formation of the 
partonic wave function, describing nuclei at the collision moment, can $not$ be adequately 
 described by the dual gravity. We know from experiment that growing partonic density
 makes hadrons and nuclei blacker and of larger size, as the collision energy grows.
 This is usually described by a Pomeron fit in which cross section $\sim s^{\alpha(t)-1}$.
 Although qualitatively similar to what happens in gravitational collisions,
 this growth is very  compared to that predicted in dual gravity. Indeed,
 the observed Pomeron intercept $\alpha(t=0)-1\sim 0.1$ while 
 in the AdS/CFT world the Pomeron intercept  $\alpha(t=0)-1=1$\cite{Polchinski:2001tt}
. Thus the effective size of objects in gravitational
 collision grows with energy with an exponent ten times that in the real QCD. 
In view of this, one should clearly  give up the idea to tune  the scale $L$ or $z_0$ of the bulk colliding object to the size of the nucleus, and tune it perhaps to the parton density (``saturation scale'' $Q_s$ in the
 ``color glass'' models) of the corresponding nuclear wave function.
 
More generally, we are dealing with a complicated problem in QCD, in which the
effective coupling runs, from higher scale to lower as the collisions progress from
initial violent partonic stage toward equilibration, expansion and cooling. So in principle,
it would be logic to switch -- as smoothly as possible --
 from the weak-coupling based methods (such as classical
Yang-Mills)  to strong coupling ones (such as AdS/CFT) at certain proper time $\tau_{switch}$ appropriately
chosen by the evolution of the coupling\footnote{The so called AdS/QCD
  approach 
(see e.g.\cite{Gursoy:2007cb,Gursoy:2007er}) 
tries to incorporate the running coupling into the gravitational
framework. 
A particularly simple
example of that is a  jump of the coupling at certain ``domain wall''
scale proposed in \cite{Shuryak:2007uq}.
}.     

 Therefore one should not try to tune the parameters of the gravitational collision
model neither to initial nuclei, at $\tau=-\infty$, nor to ``decoherent'' partons at the
 collision moment, at $\tau=0$, but  at  the later time  
 $\tau_{switch}$.\footnote{It is proposed in \cite{kovchegov} that one may choose
to collide some special unphysical shock waves}
Although we at the moment do not understand the evolution of appropriate
 coupling quantitatively enough, one may always treat it as a parameter. The practical utility of the AdS/CFT
 approach at later time $\tau>\tau_{switch}$ still remains significant: namely one can use
 much more fundamental dual gravity description instead of its near-equilibrium
 approximation, the hydrodynamics, currently used.

\begin{figure}[t]
\includegraphics[width=0.55\textwidth]{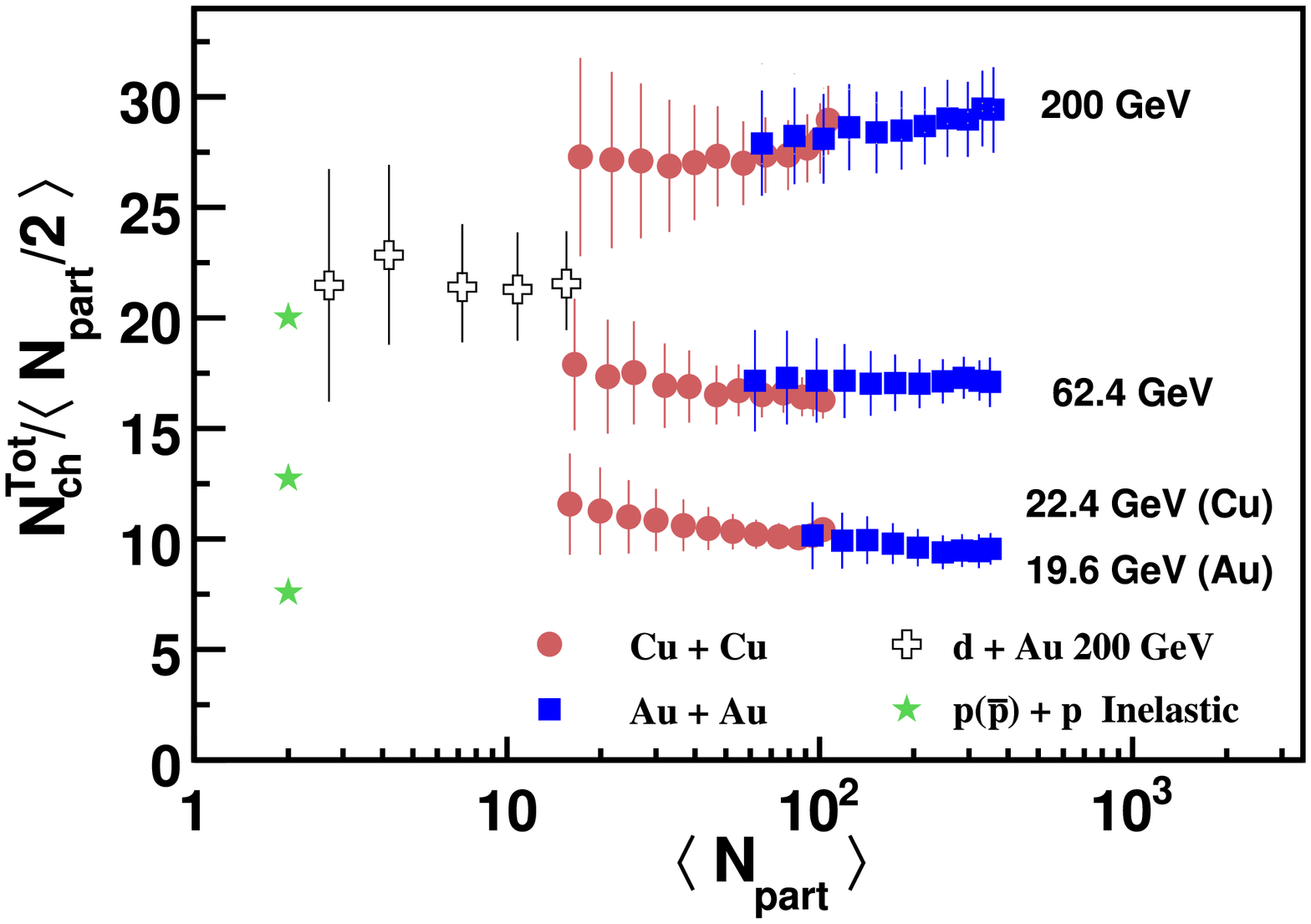}
\includegraphics[width=0.4\textwidth]{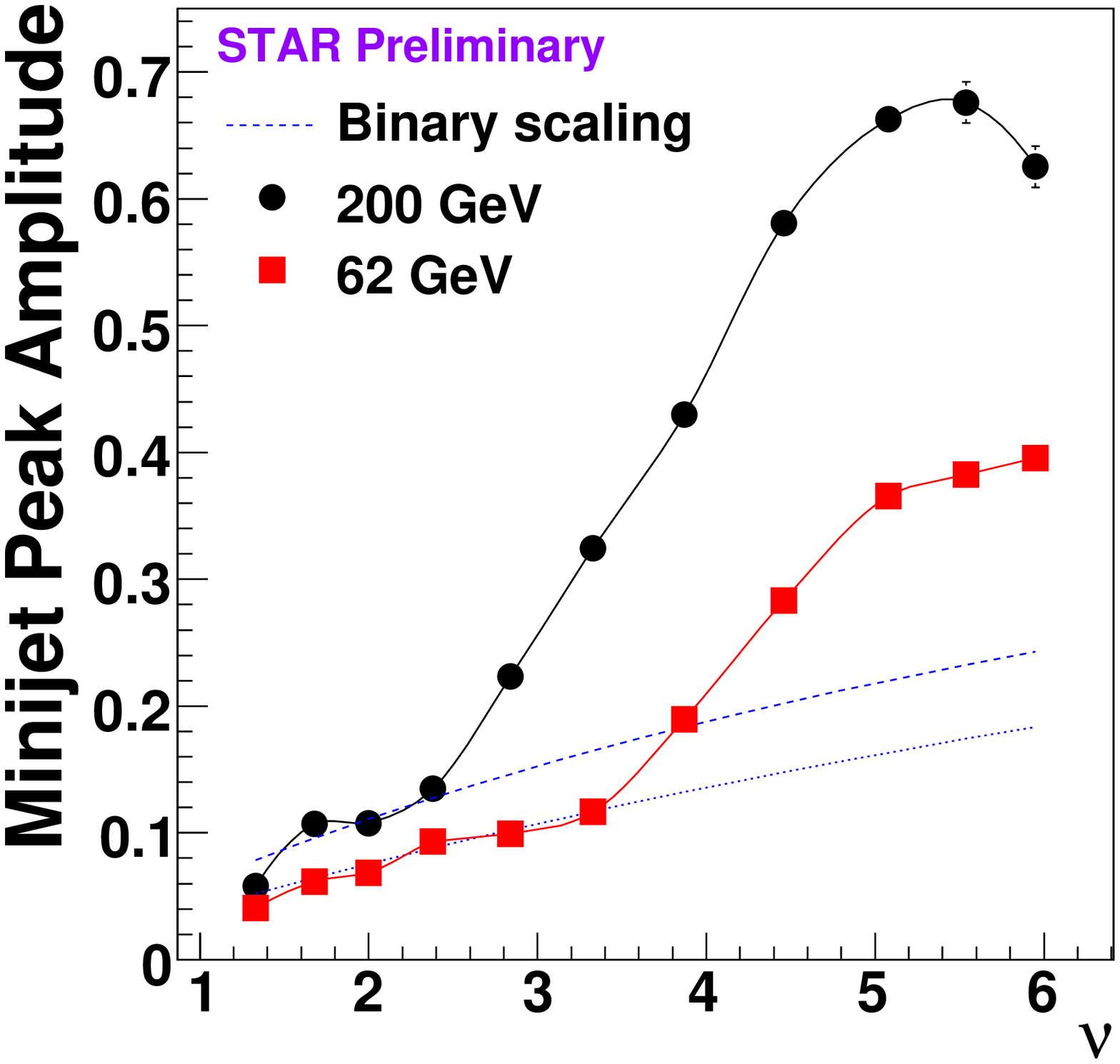}
\caption{\label{fig2} 
(left) PHOBOS data \cite{phobos} on integrated number of charged particles, scaled by $N_{part}/2$, in p+p, d+Au, Cu+Cu and
Au+Au collisions as a function of centrality. The uncertainty of Npart has been included in the error bars.
(right) The height of the ridge as a
function of the number of mean binary collisions per
nucleon. The data are from STAR collaboration  \cite{star_ridge,raju}
at two collision energies shown in the figure.
}
\end{figure}

\section{Are there critical impact parameters in heavy ion collisions?}

Summarizing our findings  in one sentence,
those is the existence of the discontinuity in grazing
gravitational collisions in the AdS space.  
As one smoothly  increases the
 impact parameter $b$, the trapped surface and black hole formation
 disappear suddenly,
at certain critical impact parameter $b_c(E)$ depending
the collision energy $E$.
The reason for this seems quite general:
 increasing $b$ one increases the angular momentum
of the system while at the same time $decrease$ the mass
which can be stopped, and at some moment -- as one knows from Kerr
solution for rotating black holes --  black hole formation
becomes impossible.

Suppose the AdS gravitational shock waves can describe
the strongly coupled plasma in heavy ion collisions:
 then one would expect similar behavior in heavy ion experiments.
We have looked at the data and found that indeed there are
experimental indications that relatively rapid switch
of the underlying dynamics at some
$b_c(E)$ seem to exist.

The most straightforward observable is entropy, related to
the particle multiplicity versus the impact parameter.
 In Fig.\ref{fig2}(left) from
\cite{phobos} we show some data plotted as a function of the
 number
of participants $N_{part}$. The right end of the figure corresponds to
all nucleons participating, or central collisions: toward the left end 
are  peripheral collisions. There are indeed
two values of multiplicity per participant observed,
 one for small systems, pp and dAu collisions (stars and crosses)
and one for ``large'' systems, CuCu and AuAu (circles and squares).
 There must be
a transition between them somewhere, but, unfortunately,
the experimental multiplicity measurements for ``grazing'' collisions 
are not available yet
\footnote{Small multiplicity collisions
are detected for all systems, but their accurate separation from beam-residual gas collisions has
not yet been systematically resolved.}. So, unfortunately,
 we do not yet know how
exactly transition from one regime to another happens and what
is $b_c(E)$, if it can be defined.

However some other observables associated with collective flows
of excited matter do show rapid changes 
at certain  $b_c(E)$ seems to be there.
 Some evidence for that were seen in the
elliptic flow measurements, as deviations from
the hydrodynamical predictions for very peripheral collisions.  
Even more clearly those are seen in the centrality dependence of the
so called ``ridge'' phenomenon (see its relation to flow
in \cite{Shuryak_ridge,raju}) which we show in Fig.\ref{fig2}(right). 

Admittedly, these
 rapid change of the dynamics have not been systematically studied yet,
neither experimentally nor theoretically. The naive
explanation often given to it attribute the change to
the fact that it happens when overlap system gets ``too small''
in terms of participating nucleons $N_p$, causing large
enough fluctuations $O(1/\sqrt{N_p})$. However, if this would be the
reason, one would expect this jump to be dependent on $N_p$
and $independent$ on the collision energy. 
Furthermore, the gravitational collisions do not have any discrete
elements at all, while predicting $b_c(E)$ growing with $E$,
as observed in Fig.\ref{fig2}. We therefore suggest
that angular momentum may also be important: this issue 
clearly deserves to be studied
further.



\section{Conclusions}

In this work  we have developed a method to solve for the shape of the trapped
surface based on an analogy to electrostatic problem in flat space: its main idea is
to proceed from differential to integral form of the equation.  We
used the method to obtain the shape of trapped surface at different impact
parameters and collision energy. We observe a critical impact parameter
within the range of energy we explored. The phenomenon is analogous to the
the critical behavior found in flat space\cite{giddings,nambu,veneziano},
the difference being the critical trapped surface depends both on
the collision energy and the nucleus size. We found
 the dependence is approximately given by a power law. Furthermore,
the shape of the critical trapped surface gets elongated in 
spherical coordinate
as the collision energy grows. We also discussed in the preceding
subsection that grazing heavy ion collisions also seem to
suggest a rapid switch  to another dynamics, without equilibration.
The exact
cause of this jump is  to be clarified in further studies.

We also studies wall-to-wall collision of 
shock waves as a simple version of the problem. The wall is
sourced by a delta function at certain initial scale $z_0$. 
We believe it is 
more reasonable initial conditions than those used by Grumiller and Romatschke 
\cite{Grumiller:2008va}, to be used in future
following their method to study  the initial stage. 

The applicability and limitation of this approach is discussed. We claim
it is more realistic to adopt partonic picture in initial stage and only
switch to effective gravity treatment at some time $after$ 
collision, when the coupling
becomes strong enough. However, we argue that the observed critical 
phenomenon is still relevant for heavy ion collisions, where
there also seems to be rapid change of collision regime as a function
of impact parameter.

Finally we would like to mention very recent work by Alvarez-Gaume et al\cite{spain}  who
discussed another extension of the problem. They considered 
 central collision of shock waves sourced by certain nontrivial matter
distribution in the transverse space. They in particularly discuss critical 
phenomenon occurring as the shock
wave reaches some diluteness limit and the formation of the 
trapped surface is no longer possible.
 It would  obviously be interesting to study how the 
two forms of critical phenomenon are related.

{\it Note added in version 2}.
 Horatiu Nastase 
had informed us about his early work \cite{Kang:2004jd}
in which he also addressed non-central
 gravitational collisions, see also
summary in \cite{Nastase:2008hw}. His estimate
of the critical impact parameter
 $b_c(E)\sim E^{1/6}$ at large $E$ is quite different
from our power, perhapse because we are not working
 at asymptotically large $E$.  We were also 
provided by (so far not posted) work by the Princeton group
\cite{Gubser_new}, who compared our numerical
results for noncentral collisions with their analytical
formulae and observed good agreement.


\noindent{\large \bf Acknowledgments} \vskip .35cm We thank
Silviu Pufu, Kevin Dusling and Stanislav Srednyak for valuable discussions. 
S.L. would like to thank Tom Kuo and
Huan Dong for help on integral equation. Our
 work was partially
supported by the US-DOE grants DE-FG02-88ER40388 and
DE-FG03-97ER4014.

\vskip 1cm


\begin{thebibliography}{99}

\bibitem{SZ12}   E.V.Shuryak,
  Prog.\ Part.\ Nucl.\ Phys.\  {\bf 53}, 273 (2004)
  [ hep-ph/0312227].
E.V.Shuryak and I. Zahed, {\tt hep-ph/0307267},
Phys.\ Rev.\ C {\bf 70}, 021901 (2004)
Phys.\ Rev.\  {\bf D69} (2004) 014011.
[ hep-th/0308073].

\bibitem{adscft}  J.~M.~Maldacena,
  Adv.\ Theor.\ Math.\ Phys.\  {\bf 2}, 231 (1998)
  [Int.\ J.\ Theor.\ Phys.\  {\bf 38}, 1113 (1999)]
  [arXiv:hep-th/9711200].



\bibitem{Grumiller:2008va}
  D.~Grumiller and P.~Romatschke,
  JHEP {\bf 0808}, 027 (2008)
  [arXiv:0803.3226 [hep-th]].
\bibitem{Gubser:2008pc}
  S.~S.~Gubser, S.~S.~Pufu and A.~Yarom,
  Phys.\ Rev.\  D {\bf 78}, 066014 (2008)
  [arXiv:0805.1551 [hep-th]].

\bibitem{kovchegov}
  J.~L.~Albacete, Y.~V.~Kovchegov and A.~Taliotis,
  JHEP {\bf 0807}, 100 (2008)
  [arXiv:0805.2927 [hep-th]].

\bibitem{witten1}E.Witten, Adv.Theor.Math.Phys.2,235 (1998), hep-th/9802150.
\bibitem{thermo} S.S.Gubser, I.R.Klebanov and A.A. Tseytlin, {Nucl.\ Phys.\ } {\bf B534} (1998) 202
\bibitem{Son}G.~Policastro, D.~T.~Son and A.~O.~Starinets,
Phys.\ Rev.\ Lett.\  {\bf 87} (2001) 081601.
\bibitem{CT}J.~Casalderrey-Solana and D.~Teaney,
 {{\tt hep-ph/0605199}}.
\bibitem{jet}S.-J. Sin and I.~Zahed,
 {\em
  Phys. Lett.} {\bf B608} (2005) 265--273,
  {{\tt hep-th/0407215}}.\\
H.~Liu, K.~Rajagopal, and U.~A. Wiedemann,
  hep-ph/0605178.\\
C.~P. Herzog, A.~Karch, P.~Kovtun, C.~Kozcaz, and L.~G. Yaffe, 
{{\tt hep-th/0605158}}.\\
S.~S. Gubser, 
A.~Buchel, 
 {{\tt
  hep-th/0605178}}.
 {{\tt hep-th/0605182}}.\\
S.-J. Sin and I.~Zahed, 
  {{\tt hep-ph/0606049}}.

\bibitem{Shuryak:2008eq}
  E.~Shuryak,
  Prog.\ Part.\ Nucl.\ Phys.\  {\bf 62}, 48 (2009)
  [arXiv:0807.3033 [hep-ph]].


\bibitem{giddings}
  D.~M.~Eardley and S.~B.~Giddings,
  Phys.\ Rev.\  D {\bf 66}, 044011 (2002)
  [arXiv:gr-qc/0201034].

\bibitem{nambu}
  H.~Yoshino and Y.~Nambu,
  Phys.\ Rev.\  D {\bf 67}, 024009 (2003)
  [arXiv:gr-qc/0209003].

\bibitem{veneziano}
  E.~Kohlprath and G.~Veneziano,
  JHEP {\bf 0206}, 057 (2002)
  [arXiv:gr-qc/0203093].

\bibitem{hawking}
  S.~W.~Hawking and R.~Penrose,
  Proc.\ Roy.\ Soc.\ Lond.\  A {\bf 314}, 529 (1970).

\bibitem{maldacena}
  D.~E.~Berenstein, R.~Corrado, W.~Fischler and J.~M.~Maldacena,
   ``The operator product expansion for Wilson loops and surfaces in the  large
  Phys.\ Rev.\  D {\bf 59}, 105023 (1999)
  [arXiv:hep-th/9809188].

\bibitem{danielsson}
  U.~H.~Danielsson, E.~Keski-Vakkuri and M.~Kruczenski,
  JHEP {\bf 9901}, 002 (1999)
  [arXiv:hep-th/9812007].

\bibitem{SSZ}  E.~Shuryak, S.~J.~Sin and I.~Zahed,
  arXiv:hep-th/0511199.
\bibitem{Janik2}  R.~A.~Janik and R.~Peschanski,
 Phys.\ Rev.\ D {\bf 73}, 045013 (2006)
  arXiv:hep-th/0512162,
  arXiv:hep-th/0606149.
\bibitem{SinNak}   S.~Nakamura and S.~J.~Sin,
  arXiv:hep-th/0607123.
   Phys.\ Lett.\ B {\bf 608}, 258 (2005)
  [arXiv:hep-th/0310031].
  
\bibitem{Heller:2008fg}
  M.~P.~Heller, R.~A.~Janik and R.~Peschanski,
  Acta Phys.\ Polon.\  B {\bf 39}, 3183 (2008)
  [arXiv:0811.3113 [hep-th]].
  
\bibitem{Lin:2008rw}
  S.~Lin and E.~Shuryak,
  Phys.\ Rev.\  D {\bf 78}, 125018 (2008)
  [arXiv:0808.0910 [hep-th]].
  
\bibitem{Chesler:2008hg}
  P.~M.~Chesler and L.~G.~Yaffe,
  arXiv:0812.2053 [hep-th].

\bibitem{Polchinski:2001tt}
  J.~Polchinski and M.~J.~Strassler,
  Phys.\ Rev.\ Lett.\  {\bf 88}, 031601 (2002)
  [arXiv:hep-th/0109174].

\bibitem{phobos}
  G.~I.~Veres {\it et al.}  [PHOBOS Collaboration],
  arXiv:0806.2803 [nucl-ex].

\bibitem{Shuryak_ridge}
  E.~V.~Shuryak,
  Phys.\ Rev.\  C {\bf 76}, 047901 (2007)
  [arXiv:0706.3531 [nucl-th]].

\bibitem{star_ridge} M.Daugherity (for the STAR coll.), Anomalous
  centrality variation..., QM08, J.Phys.G.Nucl/Part.Phys. 
35 (2008) 104090

\bibitem{raju}
  A.~Dumitru, F.~Gelis, L.~McLerran and R.~Venugopalan,
  Nucl.\ Phys.\  A {\bf 810}, 91 (2008)
  [arXiv:0804.3858 [hep-ph]].



\bibitem{spain}
  L.~Alvarez-Gaume, C.~Gomez, A.~S.~Vera, A.~Tavanfar and M.~A.~Vazquez-Mozo,
  arXiv:0811.3969 [hep-th].

\bibitem{Gursoy:2007er}
  U.~Gursoy, E.~Kiritsis and F.~Nitti,
  JHEP {\bf 0802}, 019 (2008)
  [arXiv:0707.1349 [hep-th]].

\bibitem{Gursoy:2007cb}
  U.~Gursoy and E.~Kiritsis,
  JHEP {\bf 0802}, 032 (2008)
  [arXiv:0707.1324 [hep-th]].

\bibitem{Veres:2008nq}
  G.~I.~Veres {\it et al.}  [PHOBOS Collaboration],
  arXiv:0806.2803 [nucl-ex].

\bibitem{Shuryak:2007uq}
  E.~Shuryak,
  arXiv:0711.0004 [hep-ph].

\bibitem{Nastase:2008hw}
  H.~Nastase,
  Prog.\ Theor.\ Phys.\ Suppl.\  {\bf 174}, 274 (2008)
  [arXiv:0805.3579 [hep-th]].

\bibitem{Kang:2004jd}
  K.~Kang and H.~Nastase,
  Phys.\ Rev.\  D {\bf 72}, 106003 (2005)
  [arXiv:hep-th/0410173].

\bibitem{Gubser_new}
  S.~S.~Gubser, S.~S.~Pufu and A.~Yarom,
  ``Off-center collisions in AdS5 with applications
  to multiplicity estimates in heavy-ion collisions''

\end{thebibliography}
\end{document}